\documentclass[12pt,pdfa=false]{article}

\usepackage{amsmath,amsfonts,amssymb,bm,slashed,bbm,mathrsfs}
\usepackage{amssymb}\usepackage{bm}            \usepackage{amsmath}\usepackage{color}
\usepackage{graphicx,xcolor}\usepackage{float}
\renewcommand{\Ref}[1]{\eqref{#1}}
\newcommand{\eq}[2]{\begin{align}\label{#1}#2\end{align}}

\newcommand{\tra}{\top}
\newcommand{\nn}{\nonumber}

\newcommand{\pa}{\partial}

\newcommand{\lsim}{\stackrel{\scriptstyle <}{\phantom{}_{\sim}}}

\newcommand{\ep}{\epsilon}

\newcommand{\al}{\alpha}\newcommand{\be}{\beta}

\newcommand{\la}{\lambda}
\newcommand{\om}{\omega}\newcommand{\Om}{\Omega}\newcommand{\Omt}{\tilde{\Omega}}
\newcommand{\vphi}{\varphi}

\renewcommand{\L}{{\cal L}}
\newcommand{\epp}{{\tilde\ep}}

\newcommand{\x}{\vec x}
\newcommand{\A}{\vec A}
\newcommand{\nnabla}{\vec\nabla}
\newcommand{\e}{\vec e}	
\renewcommand{\v}{\vec v}	
\newcommand{\D}{\vec D}

\begin{document}

 \title{Generation of a scalar vortex in a rotational frame}

\author{
		M. Bordag\footnote{bordag@theor.jinr.ru}\\
		\small 	
Bogoliubov Laboratory of Theoretical Physics,
Joint Institute for Nuclear Research, Joliot-Curie street 6,\\
		\small 141980 Dubna, Russia\\
	 	\small Institute for Theoretical Physics, University Leipzig,
		IPF 231101, D-04081 Leipzig, Germany
		\and
		D. N. Voskresensky\footnote{dvoskre@theor.jinr.ru} \\
		\small	
Bogoliubov Laboratory of Theoretical Physics,
Joint Institute for Nuclear Research, Joliot-Curie street 6,\\
		\small 141980 Dubna, Russia\\
		\small National Research Nuclear University (MEPhI),
		Kashirskoe shosse 31,
		115409 Moscow, Russia
	}


\maketitle

\begin{abstract}
	We consider generation from the vacuum of a scalar charged field in a rigidly rotating frame. Adding an external magnetic field opens the way to Bose condensation of the field. This phenomenon has been studied for external uniform magnetic field occupying the whole volume of the uniformly rotating cylindrical system of finite radius $R$ with a Dirichlet boundary condition imposed on it. Besides continuing this study, we consider the field formed by a flux tube of small radius. We find numerical solutions of the Gross-Pitaevskii equation for the charged scalar field, the critical rotation frequencies, the mean radii and the condensate energies, and compare them with those found in a linearization scheme and with approximate analytical solutions. We show that for the same input parameters the energy of the condensate in the case of the flux tube is lower than in the case of uniform magnetic field in the whole cylinder.
\end{abstract}


\section{\label{T1}Introduction}	
In {strong} scalar and electric external fields forming a deep potential well for pions, the vacuum becomes unstable to formation of a pion condensate \cite{migd72-34-1184}.
Pion condensation was then discussed in application to neutron stars, abnormal nuclei and heavy ion collisions, see \cite{migd78-50-107}, \cite{migd90-192-179}. We need not mention the huge field of non-relativistic applications of the Bose-Einstein condensation \cite{PitString,PS2002}. Another interesting area is rotating fields. These gained interest over the past years in connection with heavy ion collisions and the QCD-lattice calculations, \cite{cher21-103-045027}, \cite{cher18-98-065016}, \cite{brag24-110-014511}.

{As} an example, the global polarization of a $\Lambda (1116)$ hyperon observed by the STAR collaboration in non-central Au-Au collisions \cite{Adamczyk} indicated the existence of a vorticity with rotation frequency
$\Omega\simeq 10^{22}$ Hz $\simeq 0.05{m_\pi}$,
where
$m_\pi\simeq 140$\,MeV
is the pion mass (in units $\hbar=c=1$ which will be used in this paper). Since the radius of the nuclear fireball is roughly
$R\sim 10/{m_\pi}$,
the rotation speed reaches values
$\Om \sim 0.5\,/{R}$.
Another interesting feature of such experiments is the existence of strong magnetic fields of order $B\sim {m_\pi^2}$ or even larger at least at the initial stage of the relativistic heavy ion collisions, cf. \cite{Voskresensky:1980nk,IST}, and possibly in interiors of compact stars. Also, the influence of {both} rotation and magnetic field on the phase structure of QCD is under discussion \cite{cher21-103-045027,Chen:2021aiq}.

It is common to introduce rotation of the complex scalar field by a metric that follows from a Galilean-like transformation of the angular variable $\vphi\to\vphi+\Om t$ (see eq. \Ref{2.1.5} below). Since it describes a globally uniform rotation like a rigid body, one is forced to constrain space by some boundary condition on a ``mathematical'' cylinder whose artificially introduced radius $R$ and rotation frequency $\Om$ are bounded by the condition $\Om R<{1}$ to {satisfy} causality, see \cite{LP1981}.
{The} discussion of other possibilities for the metric, as suggested in \cite{fran22-8-265} (for a more recent discussion see \cite{delo99-29-1233}), is beyond the scope of this paper.

A scalar field is employed to describe neutral $\sigma$ mesons{;} a pseudoscalar field is used to describe pions.
In general, rotation, like an electric field {in the form} of a potential well, lowers the single-particle energy levels.
Unlike an electric potential well, rotation alone is not sufficient to cause a vacuum instability. In an instructive picture, this is because of the centrifugal force. {An} external electric potential well or a magnetic field may counteract the centrifugal force and further lower the energy until instability is reached. A new vacuum state, which is characterized by a Bose condensate, will emerge. A condensation of bosons must be stopped by some mechanism. One uses a $\phi^4$ self-interaction for this and electrostatic self-repulsion \cite{migd78-50-107,migd90-192-179}. The back-reaction of the induced magnetic field may also be suitable.

We mention recent studies of a possible condensation of bosons due to rotation and application of a homogeneous external magnetic field \cite{liuz18-120-032001,guot22-106-094010}. In \cite{liuz18-120-032001} an electrostatic self-repulsion was included together with restriction of the orbital moments to the number of states. In \cite{guot22-106-094010}, a $\phi^4$ self-interaction was incorporated.
{The} effect of external static electric potential well was included in \cite{vosk24-109-034030,vosk24-21-1036} and relevance of the Meissner effect and the London moment was discussed in \cite{vosk25-111-036022}.

At this {point} we should note that in a realistic treatment the equations for the scalar charged field, the magnetic and electric fields under the rotation should be solved self-consistently. This is a complicated problem. {Therefore} in the present paper as in \cite{liuz18-120-032001,guot22-106-094010} we continue the investigation of the opportunity for condensation of a charged scalar or a pseudoscalar field due to rotation assuming that we deal with an external magnetic field ignoring the back influence of the condensate on the electromagnetic field. We study two cases for the magnetic field inside the cylinder: a homogeneous field in the whole volume of the cylinder (a rotating empty solenoid in a realistic problem) and a thin vortex flux tube (solenoid), whose radius $r=R_s$ is much smaller than the radius of the cylinder $R$. 
We are interested in the critical frequency $\Om_{cr}$, at which production of charged scalar bosons and condensation of the boson field begin under the action of the rotation and external magnetic field.
The input parameters in this setup are the mass $m$ and the coupling $\la$ of the self-interaction of the field, as well as the external rotation frequency $\Om$, the value of the external uniform magnetic field $B_0$, the radius $R$ of the cylinder, and the magnetic flux $\delta_\Phi=B_0 R^2/2$, as well as the radius of the thin flux tube $R_s$. Of course, the critical rotation frequency must also satisfy the causality constraint, $\Om_{cr}R<{1}$. We investigate the energetically favored state of lowest (negative) condensate energy, as a function of the mentioned input parameters. {The} radius of a thin flux tube will enter only through the value of the flux. The value of the angular momentum $l$ of the produced field is expressed via the mentioned input parameters. The condensate field can be found {by} solving the Gross-Pitaevskii-like equation. This equation is non-linear. As a method, we apply the linearization procedure and compare the results with a numerical evaluation of the full equation and with approximate analytical solutions.

The paper is organized as follows. In the next section we {recall} the basic formulas for a field in a rotating frame, the Lagrangian and a variational approach for the linearized solution of the Gross-Pitaevskii-like equation and {for} the condensate energy. In the following section we perform the basic calculation for two models of the magnetic background field{:} a thin flux line and a field {extended} to the whole cylinder. In the last section we formulate conclusions. Some formal details are deferred to the Appendices.

\smallskip


\section{\label{T2}Basic formulas and notations}
%
\subsection{\label{T21}Resting and rotating frames}
We start with the notations for four- and three-dimensional vectors. The latter are denoted by an arrow.
Coordinates and fields are taken as contravariant vectors, whereas the derivatives are covariant,
\eq{2.1.1}{ x^\mu=\left(\begin{array}{c}x_0\\ \x\end{array}\right) ,\qquad
	A^\mu=\left(\begin{array}{c}A_0\\ \A\end{array}\right) ,
	\qquad \pa_\mu=\left(\begin{array}{c}\pa_0 \\ \nabla\end{array}\right) .
}
We start with a resting frame. Here, coordinates are marked by an index '$R$'.
For the spatial part we use the cylindrical coordinates in parallel to the Cartesian ones,
\eq{2.1.3}{\x_R=r\e_r(\vphi_R),\qquad 	
	A_R^\mu=\left(\begin{array}{c}A_{R,0}(x_R)\\ \A_R(x_R)\end{array}\right),
}
and $A_R^\mu$ is some vector field.
The radius $r=\sqrt{x_1^2+x_2^2}$ will not be transformed and is left without index $R$.
In the following we are concerned only with fields where the zeroth component is a function of the radius $r$ and where the vector part takes the form
\eq{2.1.4}{ A_{R,0}(x_R)= A_{R,0} (r),\qquad
	\A_R(x_R)=\e_\vphi(\vphi_R)\frac{\mu(r)}{r},
	\qquad \pa_{R,\mu}\equiv \frac{\pa}{\pa x_R^\mu}=\left(\begin{array}{c} \pa_0 \\ \nabla_R\end{array}\right).
}
Here $ A_{R,0}(r)$ may be used to describe an electric potential and $\A(x_R)$ is the vector potential with a dimensionless profile function $\mu(r)$. 

The transition from the resting to the rotating frame is {achieved} by a transformation of the angular variable,
\eq{2.1.5}{ \varphi_R=\varphi+\Om t,\quad t_R=t, \quad r_R=r,
	\quad z_R=z=x_3,
}
where
the vector of external rotation frequency is $\vec \Om=\Om\e_z$.
The points $\x=r\e_r(\vphi)$ and $\x_R=r\e_r(\vphi_R)$ in both frames are related by
\eq{2.1.5a}{ \x = \bm{D}^\tra(\Om t)\ \x_R,\quad
	\bm{D}(\Om t)=\left(\begin{array}{ccc}\cos(\Om t)&-\sin(\Om t) & 0\\
		\sin(\Om t) & \cos(\Om t) & 0\\0&0&1\end{array}\right),
}
where $\bm{D}$ is the rotation matrix.
The speed of a point in the rotating frame is
\eq{2.1.6}{  \v=\dot \x = \Om r\e_\vphi(\vphi) =\vec\Om\times \vec{x}
	=\left(\begin{array}{c}-\Om x_2 \\ \Om x_1 \\ 0 \end{array}\right).
}
The transformation \Ref{2.1.5} results in a rigidly rotating frame. The speed of light restricts its applicability {to} $\Om\, r\le 1$.

From the transformation between the frames one derives the transformation matrix,
\eq{2.1.7}{ {T^\mu}_\nu = \frac{\pa x_R^\mu}{\pa x^\nu} = {\left(\begin{array}{cc} 1  & 0\\-\v_R  
			  & \bm{D}(\Om t)		\end{array}\right)^\mu}_\nu,
}
the interval and the metric,
\eq{2.1.8}{ ds^2 &=
	\eta_{\mu\nu}dx_R^\mu dx_R^\nu=g_{\mu\nu}dx^\mu dx^\nu,\qquad
	\\\nn		g_{\mu\nu}&={T_\mu}^{\mu'}\, {T_\nu}^{\nu'}\eta_{\mu'\nu'}
	= \left(\begin{array}{cc}1-\v^2 & -\v^\tra\\
		-\v&-\bm{1}\end{array}\right),\quad
	g^{\mu\nu}= \left(\begin{array}{cc}1 & -\v^\tra\\
		-\v&-\bm{1}+\v\circ  \v^\tra\end{array}\right).
}	
We mention that the metric \Ref{2.1.8} is flat; all curvature tensors are zero and $\det g=-1$ holds.

In the rotating frame
\eq{2.1.9}{\pa_\mu={T_\mu}^\nu \ \pa_{R,\nu}=
	\left(\begin{array}{c}\pa_0-\v\,\nabla \\ \nabla \end{array}\right),
	\quad\pa^\mu={T^\mu}_\nu\ \pa_R^\nu=\left(\begin{array}{c}\pa_0 \\ -\nabla \end{array}\right),
}
holds for the derivatives and
\eq{2.1.10}{
	A^\mu(x)& ={T^\mu}_\nu \ A_R^\nu(x_R)= \left(\begin{array}{c}A_{R,0}(x) \\ -\v\, A_{R,0}(x)+\A_R(x) \end{array}\right),
	\\\nn	A_\mu(x)&=\left(\begin{array}{c} A_{R,0}(x) -\v\,\A_R(x) \\ -\A_R(x) \end{array}\right),
}
for the field.
Indices are raised and lowered using the metric, as usual.

\subsection{\label{T22}Lagrangian and field equation in the rotation{al} frame}
The Lagrangian density for a charged scalar or pseudoscalar field in a curved background is
\eq{2.2.1}{ \L = \sqrt{-g}\left[
	g^{\mu\nu}\left(D_\mu\phi\right)^* D_\nu\phi
	-(m^2+\xi R)\phi^*\phi-\frac{\la}{2}(\phi^*\phi)^2\right],
}
with the mass $m$, the coupling $\xi$ to the scalar curvature $R$ and the coupling constant $\la>0$ and the 'long' derivative,
\eq{2.2.2}{D_\mu=\pa_\mu-i e A_\mu,
}
where $-e$ is the charge of the electron, $e^2\approx 1/137$, $\pa_\mu$ in the {rotational} frame is determined in \Ref{2.1.9}. Although the most physically meaningful {case is that of the} pseudoscalar charged pion field, for definiteness let us further speak about the scalar field.
In the rotation{al} frame it is a function of coordinates, $\phi(t,\x)$, which we suppress temporarily for simplicity of the formulas.
The derivatives are partial ones for the time-space part (in the Lorentz indices); since $\phi$ is a scalar field no Christoffel symbols enter. Below, since the metric \Ref{2.1.8} is flat, we need neither $\sqrt{-g}$ nor the curvature $R$.
We consider the field inside a cylinder of some radius $R$, fulfilling the causality condition $\Om R\le 1$ and must thereby impose the boundary condition. We take the Dirichlet condition,
$\phi(\x)=0$ at $r=R$. With this condition, surface terms will not appear when integrating by parts in the Lagrangian.
Now, the Lagrangian density \Ref{2.2.1} can be rewritten as
\eq{2.2.3}{ \L=\phi^* \left[-D_\mu g^{\mu\nu}D_\nu-m^2-\frac{\la}{2}\phi^*\phi
	\right]\phi\,.
}
From the minimum of the action corresponding to the Lagrangian density \Ref{2.2.3} one derives the Euler-Lagrange equation (Klein-Gordon equation with interaction),
\eq{2.2.4}{  \left(- (D_0-\v\,\D )^2+\D^2-m^2-{\la}\phi^*\phi\right)\phi=0,
}
and its complex conjugate. This equation, the GP-like equation for short, is a relativistic generalization of the Gross-Pitaevskii equation for the order parameter. In our case the classical field $\phi$ appearing at some conditions from the vacuum as we will see plays the role of the order parameter. The field $A_\mu$ in this work we treat as an external field, as we have mentioned.

In the following, dealing with a static magnetic background field $\vec{A}_R$ we put $A_{R,0}=0$ in \Ref{2.1.3} and \Ref{2.1.10}. Then we are able to focus on the case of the stationary condensate field $\phi(t,\x)\propto e^{-i\om t}\phi(\x)$ occurring at the corresponding conditions in the {rotational} frame. The energy density corresponding to the Lagrangian density is ${\cal{E}}=\om n -\L$, where
$n=\partial \L/\partial \om$
is the particle density.
With increasing rotation speed $\Om$, the ground state energy of the boson $\om$ first reaches zero, $\om =0$, and then {dives} further {down} to $-m$. Then in the passive rotation problem bosons of one sign {of} charge can be produced occupying the level $\om=-m$ in a tunneling process from the lower to the upper continuum and anti-bosons go off to infinity. To be careful, note that in a short-range electric potential well at $mR\gg 1$ the anti-particle energy level appears near the boundary of the lower continuum, at $\om\simeq -m+O(1/R)$, see \cite{migd72-34-1184}. In the presence of the empty rotating ``physical'' cylinder, instability may occur already when the ground state single-particle energy level reaches zero{;} then the reactions with production of scalar particles and antiparticles on the surface of the cylinder become possible. Antiparticles are absorbed by the surface of the cylinder. Particles form the classical field $\phi (\vec{x})$ inside the cylinder. This mechanism of production of the field $\phi$ was discussed in \cite{vosk24-109-034030,vosk24-21-1036}.
For $\om>0$ the number of particles, $n$, calculated within the single-particle problem is non-negative, see \cite{vosk25-111-036022}. {The} minimum of $\om n$ corresponds to $\om =0$.
In this case when the single-particle ground state energy level, $\om$, reaches zero a new vacuum state can be occupied, characterized by the classical static ($\om =0$) field $\phi$, which is described by the GP-like equation. The value
$\om=0$ plays the role of the chemical potential for the order parameter $\phi$, see \cite{migd78-50-107}, \cite{migd90-192-179}.
Equation \Ref{2.2.4} turns then into
\eq{2.2.6}{\left[-(\v\,\nnabla)^2+\D^2-m^2
	-{\la}\phi(\x)^*\phi(\x)\right]\phi(\x)=0.
}
The energy per unit (longitudinal) length of the classical field $\phi(\x)$ in the cylinder follows from the Lagrangian density (\ref{2.2.3}) and reads
\eq{2.2.7}{ E_{GP} = \int d^2x \left[-|(\v\,\nnabla)\phi)|^2
	+|\D\phi|^2+m^2|\phi|^2+\frac{\la}{2}|\phi|^4\right],
}
or, {using} eq. \Ref{2.2.6} and integrating by parts,
\eq{2.2.8}{E_{GP} =- \frac{\la}{2} \int d^2 x \  |\phi|^4.
}

The GP-like equation \Ref{2.2.6} is nonlinear and does not have an exact analytical solution. So one is left with approximate analytical and numerical ones. As one possibility, we use a linearization as an approximation and check the results by numerical evaluation. In eq. \Ref{2.2.6} we substitute the non-linear term by a constant,
\eq{2.4.1}{\la\phi^*(\x)\phi(\x) \to \epp^2.
}
Here we follow \cite{bord25-40-2543018}, where this method was discussed recently.
It was shown that this approximation may be considered as a zeroth step in a perturbative expansion in powers of $\epp$ as an expansion parameter.

After substitution \Ref{2.4.1} into \Ref{2.2.6}, the linearized equation becomes
\eq{2.4.2}{ \left( (\v\,\nnabla)^2-\D^2+m^2 \right) \phi=-\epp^2 \phi.
}
Formally, this looks like a Klein-Gordon equation. The ground state solution with real $\epp=\epp_0$ describes a state with the condensate. We mention that at zero temperature, which we consider here, only the lowest level will be populated by the condensate.


\section{\label{T3}Magnetic and scalar fields in the rotating frame}

\subsection{\label{T23}Notations for the background field}
In this paper we restrict ourselves {to studying} the response of the charged scalar field vacuum {to} application of the external magnetic field of a specific form in the resting frame.
As a background external field we consider a cylindrically-symmetric magnetic vector potential $\A_R(x_R)$ of a solenoid ({hereafter} flux tube). The relation between variables and fields in the resting and {rotational} frames is given by eq. \Ref{2.1.10}. In the rotating frame in variables of the {rotational} frame $\A(x)$ has the same profile function $\mu(r)$ as in the resting frame,
\eq{2.3.1}{ \A(\x)=\e_\vphi\frac{\mu(r)}{r},
	\qquad   \mu(r)=\frac{B_0 r^2}{2}\Theta(R_s-r) + \frac{B_0 R_s^2}{2}\Theta(r-R_s),
	\qquad 0<R_s<R\,.}
Here the step-function $\Theta(x)=1$ for $x\geq 0$ and zero for $x< 0$.
It describes a homogeneous magnetic field $\vec B=B_0 \e_z$ inside a flux tube with radius $R_s$ and zero field outside,
\eq{2.3.1a}{\vec B=\nnabla\times \A = \e_z\frac{\mu'(r)}{r},\qquad
	\vec B = \e_z\, B_0 \, \Theta(R_s-r).
}
For $R_s =R$ the field \Ref{2.3.1} describes the field in the whole cylinder of {radius} $R$, $B_0=const$. For definiteness let $B_0>0$.
The corresponding classical electromagnetic energy per unit length, $E_{el}$, and the flux $\Phi\equiv 2\pi\delta_\Phi/e$ of the background field are given by \cite{LP1981},
\eq{2.3.2}{
\quad
	E_{el}=\frac{\pi}{8\pi} B_0^2 R_s^2
	\equiv  \frac{\delta_\Phi^2}{4e^2 R_s^2} ,\qquad 	\delta_\Phi=\frac{e B_0 R_s^2}{2} ,
}
where $\delta_\Phi$ is a convenient notation below in the Kummer's function.

In the   limit $R_s\to R$, the uniform magnetic field fills the whole cylinder.
In the opposite limit of small $R_s$, the magnetic field is confined to a narrow flux tube.
Keeping the flux $\Phi$ fixed, the magnetic field $B_0=\frac{\Phi}{\pi R_s^2}$ becomes large. Mathematically, in the limit $R_s\to0$, it can be defined in terms of a distribution (generalized function)   and results in
\eq{2.3.4}{ \vec B=
	\lim\limits_{R_s\to 0}\e_z\frac{\mu'(r)}{r}
	=\e_z\Phi \, \delta^2(\x_{(2)}),
}
where $\x_{(2)}=\left(\begin{array}{c}x_1\\x_2\end{array}\right)$ , and   $\delta^2(\x_{(2)})=\delta(x^1)\delta(x^2)$ is a two dimensional delta function.  This idealization is sometimes used  to describe an infinitely thin flux tube.
Since we study the case $\mu=\mu (r)$, it is the same in both the  resting and rotating frames.

As we have mentioned, in a formal mathematical problem it is sufficient to say that we consider the problem in uniformly rotating frame at $\Om R<1$.
In a realistic consideration  the rotation frame  is associated
with a rotating body; in our case with the flux tube of the radius $R_s$ (empty thin wall cylinder of radius $R_s$) placed inside the empty thin-wall cylinder of radius $R\geq R_s$, both rotating with the same external rotation frequency $\vec{\Om}\parallel z$. Bearing in mind future possible applications to heavy-ion collisions we will orient on values of input parameters $\Om$, $B_0$, $R$, $\lambda$, $m$,
mentioned in the Introduction, relevant for pions in relativistic heavy-ion collisions.

\subsection{\label{GPeqs}Gross-Pitaevskii like equation for the condensate }

We continue to consider a scalar field $\phi(t,\x)$ in the rotating frame given by the metric \Ref{2.1.8} inside a cylinder. Its radius $R$ and the rotation frequency $\Om$ are constrained by causality condition $\Om R<1$. Also we will focus on the physically meaningful case (e.g. for pions in heavy-ion collisions), $Rm\gg 1$.
We are interested in reaction of the charged scalar field vacuum on rotation in presence of the external magnetic field given by \Ref{2.3.1}.

Since the background field is static and cylindrically symmetric, we make the corresponding Fourier transforms,
\eq{3.0.1}{ \phi(\x)=e^{i l \vphi+i k_z z}\phi(r),
}
using cylindrical coordinates, after which the field $\phi(r)$ can be taken real.  The GP-like equation \Ref{2.2.6} turns into
\eq{3.0.2}{ \left( ( \Om l)^2
	+\pa_r^2+\frac{1}{r}\pa_r-\frac{(l-e\mu(r))^2}{r^2}
	-k_z^2-m^2-\la\phi^2(r)\right)\phi(r)=0,
}
where we used \Ref{2.3.1} for the background field, and with \Ref{2.1.5} and \Ref{3.0.1}, $\v\,\nnabla \phi(\x) \to i\Om l\phi(\x)$. In \Ref{3.0.1} we employed the translation invariance in z-direction. Since we are interested in the ground state, we put the momentum $k_z=0$.

Equation \Ref{3.0.2} must be supplemented by the boundary conditions, the regularity at the origin and Dirichlet condition at the cylinder, i.e., $\phi(R)=0$. With $\mu(r)$ from \Ref{2.3.1} one needs still boundary conditions at $r=R_s$. These are continuity of the field $\phi(r)$ and its first derivative at $r=R_s$. Example of matching of the {solutions of eq. \Ref{3.0.2} linearized following the procedure} \Ref{2.4.1}, \Ref{2.4.2}, is discussed in Appendix \ref{ApA}.
The linearized equation,
following from \Ref{2.4.2},
is as follows
\eq{3.0.3}{ \left( ( \Om l)^2
	+\pa_r^2+\frac{1}{r}\pa_r-\frac{(l-e\mu(r))^2}{r^2}
	-m^2\right)\phi(r)
	=\epp^2\phi(r).
}


\subsection{\label{T24}Energy of the linearized GP-like equation}

A solution with real $\epp$  owes its existence to the rotation term (first term in the parentheses in the above equation) and it
{exists} above a certain threshold where the   instability sets in. For the same choice of the parameters
at $\epp=0$ both equations, \Ref{2.4.2} and \Ref{2.2.6}, coincide. While the solution of the linearized equation is determined only up to a factor, the exact solution has no freedom in rescaling. Thus it will vanish for $\epp=0$. The factor of proportionality is given by eqs. (21) and (29) in \cite{bord25-40-2543018}.  So, the suggested linearization scheme  is expected to be good for small $\epp_0$.

Inserting the solution $\phi$ in eq. \Ref{2.2.7} and using eq. \Ref{2.4.2},
we express the ground state energy  of the condensate field $\phi$ per unit length as
\eq{2.4.3}{ E_{0}^{lin} =\int d^2x \left( -\epp^2_0 \al^2\phi^2
	+\frac{\la}{2}\al^4\phi^4 \right),
}
where $\al$ is a scale factor of the solution, $\phi\to \al \phi$. Although it is arbitrary since eq. \Ref{2.4.2} is linear, its value can be determined by demanding that the energy takes a minimum. One finds
\eq{2.4.4}{  \al^2=\frac{\epp^2_0}{ \la}\,\frac{a}{b},
}
where $a$ and $b$ are integrals of the function $\phi$,
\eq{2.4.5}{ a=\int d^2x \ \phi^2, \quad b=\int d^2x\  \phi^4.
}
Therefore  proper normalization of the function will be achieved by the substitution
\eq{2.4.7}{ \phi
	\to\al\phi,
}
independently of the initially taken function as long as it is used calculating $a$ and $b$.

Inserting \Ref{2.4.4} into \Ref{2.4.3}, we arrive at the result
%
\eq{2.4.8}{ E_{0}^{lin} = -\frac{\epp^4_0}{2\la}\,\frac{a^2}{b}.
}
The above scheme is a kind of variational approach and provides  with \Ref{2.4.8} an upper bound on the true energy of the condensate, \Ref{2.2.8}, which would follow from inserting of the exact solution of the non-linear equation \Ref{2.2.4} into the energy per unit length given by eq. \Ref{2.2.7}. In section \ref{T3.2} we will give an example. We mention that the coupling constant $\la$, entering equation \Ref{2.2.6},  is absorbed in \Ref{2.4.2} in the $\epp^2$-term and  reappeared as factor in the denominator  in the energy per unit length \Ref{2.4.8}.


\subsection{\label{T3.2}Rotation with a thin flux tube}
%
First we will consider the case of infinitely thin flux tube and then discuss accuracy at which this solution holds for the flux tube of finite size.
\subsubsection{\label{T3.2.1}Solution of a linearized equation}

The corresponding background field is given by eq. \Ref{2.3.4} and the profile function becomes a constant, $\mu(r)=\delta_\Phi$. The GP-like equation \Ref{3.0.2} simplifies accordingly.
In the Appendix \ref{ApA} we demonstrate that the solution of the  linearized equation \Ref{2.4.2} with a finite radius $R_s$ in the limit $R_s\to0$ turns into a Bessel function.
This
demonstrates that one can take the limit   $R_s\to 0$  directly in eq.  \Ref{3.0.2}, that is what we do.
This way, our equation for all $r{\ge}0$ becomes
\eq{3.2.1}{ \left(     -\pa_r^2-\frac{1}{r}\pa_r	+\frac{\nu^2}{r^2}
	-\Omt^2
	+\la \phi^2(r) \right)\phi(r)
	= 0,\qquad \nu=|\delta_\Phi -l|,
}
where we introduced notation
\eq{3.2.11}{
	\Omt=\sqrt{(\Om l)^2-m^2}\,.
}
We should recall that in our case as  input parameters we use the values $R,B_0,\Omega$ and for infinitely thin flux tube the values $R$ and $B_0$ enter only through the value of the magnetic flux, $\delta_\Phi$.  As we shall see below,  the value of the angular momentum $l$ in \Ref{3.2.11} corresponding to the minimal energy $E_{GP}$ is expressed via  mentioned input parameters and differs only a little from $\delta_\Phi$ and whether  in the energy minimum $l-\delta_\Phi>0$ or $<0$ is not determined since  $\nu$ enters GP-like equation \Ref{3.2.1} only as $\nu^2$.

First, we consider  eq. \Ref{3.2.1} in the approximate scheme described in section \ref{T24}. Equation \Ref{3.0.3}  turns into the equation for the  Bessel function,
\eq{3.2.2}{ \left(     -\pa_r^2-\frac{1}{r}\pa_r	+\frac{\nu^2}{r^2}
	-\Omt^2
	\right)\phi(r)
	= -\epp^2\phi(r),
}
supplemented by the boundary condition $\phi(R)=0$ and condition of regularity at $r=0$.
The magnetic field enters only through the flux in the shifted quantity $\nu$. The solution, which fulfills the boundary conditions, reads
\eq{3.2.2a}{ \phi(r)=J_\nu\left( j_{\nu,n} \frac{r}{R}\right),
}
with
\eq{3.2.3}{ \epp = \sqrt{\Omt^2
		-\left(  \frac{j_{\nu,n}}{R}  \right)^2   },
}
where $j_{\nu,n}$ are the zeros of the Bessel function. Recall that for $\nu\neq 0$ the Bessel function obeys the boundary condition $J_\nu(r\to 0)\to 0$ for $r\to 0$ and for $\nu= 0$, $J_0(r\to 0)\to const$.

In this scheme,  a solution with real $\epp$, which describes a condensate, will be precluded for large $l$ by the growth of the zeroth of the Bessel function. It is  the magnetic flux entering the index which compensates this growth. We are interested in the solution with lowest condensate energy $\epp =\epp_0 $. It is given by the lowest root, $j_{\nu,1}$, of the Bessel function  in \Ref{3.2.2a}.

From \Ref{3.2.3} it can be seen that there exists a critical (lowest) frequency $\Om_{cr}$ corresponding to  $\epp =0$,
\eq{3.2.4}{ \Om_{cr}=\frac{1}{l}\sqrt{m^2+\left(  \frac{j_{\nu,n}}{R}  \right)^2},
}
such that for $\Om>\Om_{cr}$ it is energetically favorable to produce the charged scalar field from the vacuum in the rotating frame.
For $\Om<\Om_{cr}$  there are no solutions $\phi$ with real   $\epp$ and the vacuum is stable.
Of course, the causality bound must be satisfied too, so that we get the restriction
\eq{3.2.5}{ \Om_{cr}R<\Om R<1.
}

\begin{figure}\begin{picture}(50,180)(0,0)
		\includegraphics[width=0.9\textwidth]{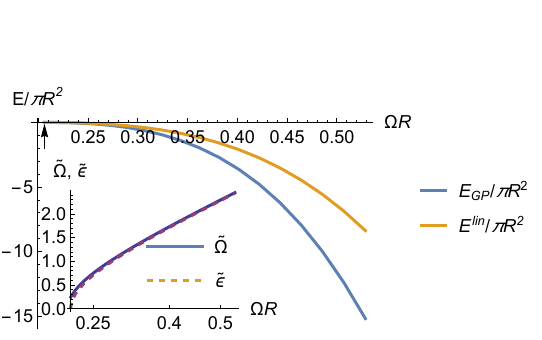}
	\end{picture}	\caption{	
		For the infinitely thin flux tube, the condensate ground state energy density  for  the exact solution,  $E_{GP}/(\pi R^2)$, see eq. \Ref{2.2.8}, and  the energy density, $E^{lin}/(\pi R^2)$, see eq. \Ref{2.4.8}, for the    solutions of the linearized equation
		as functions of the input parameter $\Om R$. Parameters $\Omt$ and $\epp$ as functions of $\Om R$ are shown in the inset. The critical value $\Om_{cr}R$ is indicated by an arrow. The parameters are $m=1$, $R=10$, $\delta_\phi=50$ and $l=50$.
	} 	\label{fig:1}	
\end{figure}

Using the formulas from section \ref{T24} and the solution \Ref{3.2.2a}  of the linearized equation one can easily calculate the condensate energy and the mean radius. As an example we take as parameters 
the magnetic field  $eB_0=m_\pi^2$ and the radius $R=10/m_\pi$, being typical values for energetic  heavy-ion collisions. Further, since we expect the lowest condensate energy for the value of the orbital momentum fully compensating the flux, we put $\nu=|\delta_\phi-l|=0$. Following \Ref{2.3.2}, this results in $l=  \delta_\phi= 50$ and is in agreement with a lower bound for the orbital moments, $l\gtrsim 10$, which follows from combining \Ref{3.2.4} with \Ref{3.2.5}.  With these parameters, the critical value of the rotation frequency  at which the condensate appears first is equal to $\Om_{cr}\simeq 0.021$. The numerical solutions confirm that this value is the same for the solution of the exact equation and the linearized one.

The  relation \Ref{3.2.5} is fulfilled even for quite large $\epp\lesssim 4.8$. The energy per unit length, $E_{0}^{lin}$, following from \Ref{2.4.8} with $\la=1$, is shown in figure \ref{fig:1} as function of  $\Omega R$. Also we show the dependence  by $\Omega R=F(\epp)$ following
\Ref {3.2.3}.
As we see from the figure \ref{fig:1}, the  energy density $E^{lin}/(\pi R^2)$ coincides well with the exact one for $\epp\lsim 1$ that corresponds to  $\Omega R\lsim 0.29$, and  as expected, in the whole region we have $E^{lin}>E_{0,GP}$. The  first excited state  solution arises from the root $j_{\nu,2}$ in the solution \Ref{3.2.2a}. It appears with increasing $\Om$ only at $\Om R\simeq 0.56$, $\epp\simeq 2.61$, and therefore it is not shown in figure \ref{fig:1}.  In the inset we show parameters $\Om$ and $\epp$ as functions of $\Om R$. They are very close to each other except the vicinity of the critical point.


Next, we estimate the validity of the linearization procedure comparing the approximate solution of \Ref{3.2.2} with the solution of the full GP-like equation \Ref{3.2.1}.
First of all we mention that for $\epp\to 0$ the full equation turns into the linearized one. Thereby, the exact solution turns into a Bessel function for $\epp\to 0$, of course up to a scale factor $\epp\frac{a}{\la b}$ with $a$ and $b$ given by \Ref{2.4.5}, calculated in \cite{bord25-40-2543018}.  So it goes to zero in the limit $\epp\to 0$ since for $\Om<\Om_{cr}$ there is no
solution with real $\epp$.

\subsubsection{\label{T3.2.2} Numerical solution of  full GP-like equation}

Here we estimate the validity of the linearization by a numerical solution of eq. \Ref{3.2.1}.
Recall that for $\epp\to 0$ the full equation turns into the linearized one. Also, the exact solution turns into a Bessel function, of course up to a scale factor, which was calculated in \cite{bord25-40-2543018}, $\epp\frac{a}{\la b}$ with $a$ and $b$ given by \Ref{2.4.5}. So the exact solution goes to zero in the limit. One may also remind that for $\Om<\Om_{cr}$ there is no real solution. The numerical solution was calculated from  eq. \Ref{3.2.1} for $\nu=0$ applying a 4th order Runge-Kutta scheme, starting in $r=0$, with  a division of the interval into $10^5$ points and application of the bisection method.

It is convenient to rewrite eq. \Ref{3.2.1} in the dimensionless variables,
\eq{3.2.9}{ \left(\partial^2_y +\frac{1}{y}\partial_y -\frac{\nu^2}{y^2}\right)\chi+\chi -\chi^3=0\,,
}
with
\eq{3.2.8}{\phi(r)=\Omt \,\frac{\chi(y)}{\sqrt{\lambda}}\,, \quad y=\Omt r\,.
}
Then the  boundary conditions render $\chi (0) =1$ and $\chi (\Omt R)=0$.
The only dimensional parameter,  $\Omt R$, appeared in the boundary condition.

%
%
The value
\eq{3.2.10}{\Omt=\sqrt{(\Om l)^2-m^2}=\sqrt{\left(\frac{j_{\nu,1}}{R}\right)^2+\epp^2}\,,
}
see eqs. \Ref{3.2.11}, \Ref{3.2.3}, characterizes variation of the amplitude of the field with the rotation frequency $\Omega$ and the angular momentum $l$, which for the ground state  coincides with  the flux $\delta_\Phi$, as we will show below.

The results for the function $\phi(r)$, \Ref{3.2.8}
are shown in figure \ref{fig:2} for several values of $\Omt$. The energy density $E_{GP}/(\pi R^2)$, see \Ref{2.2.8} for the energy per length,
is shown in  figure \ref{fig:1}.

\subsubsection{\label{T3.2.3}Approximate analytical solution of the GP-like equation}

{\bf{Case $\nu =0$.}} In the case $\nu =0$, all parameters disappear from eq. \Ref{3.2.9}. {The} typical values of the dimensionless variable $y$, at which the function $\chi$ {changes} essentially, are $y\sim 1$. This means that for $y\gg 1$ we may neglect the second (curvature) term in \Ref{3.2.9} and the problem is reduced to an effectively one-dimensional problem. Then (with the neglected curvature term) we are able to easily find the exact analytical solution, $\chi^{(1)}(y)$, of \Ref{3.2.9} satisfying boundary conditions $\chi (0) =1$ and $\chi (\Omt R)=0${;} see \cite{Migdal:1977rn} where a similar procedure was employed, although in another physical problem. The solution {is}
%
\eq{3.2.12}{ \chi^{(1)}(y)= \tanh\left(\frac{\Omt R-y}{\sqrt{2}}\right).
}
Returning to the case of the cylindrical geometry, the solution coincides with \Ref{3.2.12} in the whole interval $1\ll y<\Omt R$ for $\Omt R\gg 1$. For $y\lsim 1$ it becomes an interpolation solution satisfying the correct boundary condition $\chi (0)=1$. Even for $y\lsim 1$ there is only a minor difference {from the} exact numerical solution.

In dimensional variables this approximate solution reads
{
\eq{3.2.121}{ \phi_{appr}(r)=\frac{\Omt}{\sqrt{\lambda}} \tanh\left(\frac{R-r}{\sqrt{2}}\tilde\Om\right).
}

The numerical approximate solution of the linear equation, $\phi_{lin} (r)$, the exact numerical solution of the GP-like equation, $\phi (r)$, as well as the analytical approximate solution, $\phi_{appr} (r)$, are shown in figure \ref{fig:2}. Parameters are chosen {to be} the same as in figure \ref{fig:1}.
It can be seen that the difference between the analytical solutions and the exact ones diminishes rapidly with increasing $\Om R$. For $\Om R \geq 0.23$ these solutions are already almost indistinguishable for $r<0.5 R$ and for $\Om R \geq 0.29$ in the whole interval $0<r<R$.
With increasing $\Omt$, a plateau forms, which finally will occupy nearly the whole interval from $r=0$ {to} $r=R$.
The height of $\phi$ approaches the amplitude $\tilde\Om/\sqrt{\lambda}$ for large $\Om R$ exponentially fast. Also, the bold dots in figure \ref{fig:2} {show} values of the mean radii
\eq{3.3.14}{ \langle r\rangle =\frac{\int dx \,r\, \phi(\x)^2}
	{\int dx \,\phi(\x)^2}.
}
The value $\langle r\rangle$ increases with $\Omt$ following the peak of the function $r\phi(r)^2$. As we see, in the case of the infinitely thin flux tube the mean radius changes in a rather narrow interval, $0.5 <\langle r\rangle/R<0.7$, since the solution $\phi (r)$ is smooth and becomes closer to a step-function with increasing $\Om R$.

\begin{figure}
	\includegraphics[width=0.75\textwidth]{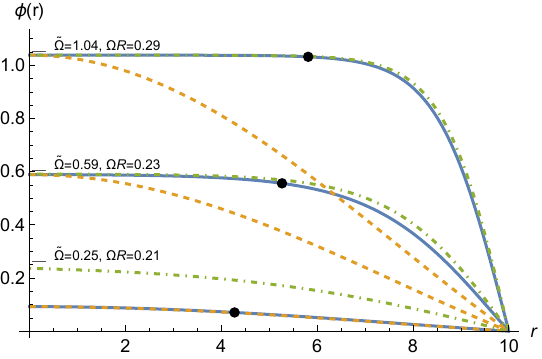}
	\caption{
		For the thin flux tube, at $\nu=0$, the exact solutions $\phi(r)$ of the GP-like equation (solid lines) and the solutions of the linearized equation (dashed lines) for several values of the parameter $\Om R$,
		$l=\delta_\Phi=50$, and $R=10, m=1, \lambda =1$. {The} dash-dotted line shows the approximate solution \Ref{3.2.121} of the GP-like equation. The dots indicate the mean radius, \Ref{3.3.14}.
	}		\label{fig:2}	
\end{figure}

As in the case of $E^{lin}$, the coupling constant $\la$, which disappeared in the dimensionless equation \Ref{3.2.9}, reappears in expression \Ref{2.2.8}, which now reads
\eq{3.2.11a}{  E_{GP} = -\frac{\pi R^2\Omt^4}{\la} \int_0^{\Omt R} dy\,y \ \chi^4(y)\,.
}

Setting the approximate solution \Ref{3.2.12} in \Ref{3.2.11a} we find the ground state energy of the field for $\nu =0$ and $\Omt R\gg1$,
\eq{3.2.12b}{
	{\cal{E}}_{GP}^{appr}\simeq {\cal{E}}^{vol}_{GP}+{\cal{E}}^{surf}_{GP}\,,\,\,
	{\cal{E}}^{vol}_{GP}\simeq
	-V \frac{\Omt^4}{{{2}}\lambda}\Theta (\Omt^2),}
where $V=\pi R^2 L_z$ is the volume of the cylinder, ${\cal{E}}^{vol}_{GP}$ is the volume contribution for $\Omt R\gg1$, and ${\cal{E}}_{GP}^{surf}\lsim O[1/(\Omt R)]$ is the surface correction term. This surface correction appears due to the deviation of $\chi (y)$ from unity at distances near the cylinder surface, $|y-R\Omt|\sim 1$. {The} region $y\lsim 1$, where \Ref{3.2.121} becomes an extrapolating solution satisfying the correct boundary condition, yields the correction to the energy of the higher order, $O[1/(\Omt R)^2]$.
With $\phi$ from \Ref{3.2.121} the surface term is given by
\eq{3.2.13b}{ {\cal{E}}^{surf}_{GP}\simeq -{\cal{E}}^{vol}_{GP}8\sqrt{2}/(3\Omt R)>0\,.}
To arrive at this result we used that $\int_0^{y_0} y\mbox{th}^4 y dy\simeq y_0^2/2 -4y_0/3$ for $y_0\gg 1$.

The value ${\cal{E}}^{lin}\propto \epp_0^4\propto (\Omt^2-\Om_{cr}^2)^2$, see \Ref{2.4.8}, \Ref{3.2.11} and \Ref{3.2.3}, demonstrates behavior typical for the second-order phase transition in the mean-field approximation which we employ, since in the critical point, $\Omt_{cr} =j_{0,1}/R\neq 0$, only the second derivative of ${\cal{E}}^{lin}$ {with respect to} $\Omega$ {has} a jump. The exact numerical solution demonstrates similar behavior. The value ${\cal{E}}_{GP}^{appr}$ given by eq. \Ref{3.2.12b} also {indicates} the second-order phase transition but at a slightly shifted value of the critical point given by $\Omt =0$.
This is because to derive the approximate analytical solution we neglected the curvature term in the GP-like equation that holds for $mR\gg 1$ and $\Omt R\gg 1$. The correction $j_{0,1}/R\sim 1/R$ is precisely of this kind since $\Omt_{cr}R\sim 1$. 

The energy density of the condensate is shown in figure \ref{fig:3} as a function of $\nu$ for the solution of the linear equation and {for} the numerical solution of the GP-like equation. The method used for the latter is the same as in section \ref{3.3.3}, see the paragraph below equation \Ref{3.3.30}. As it should be, the ground state energy per unit length, $E_0^{lin}$, \Ref{2.4.8}, calculated from the linearized equation, is above the exact value $E_{GP}$. 
The minimum of $E_{GP}(\nu)$ corresponds to $\nu =0$, as it was guessed above. The minimum of $E_0^{lin}$ is slightly shifted and corresponds to $\nu\simeq 1$ in the given example. The minimum of the energy calculated for $\nu=0$ with the analytical solution for $\Omt R\gg 1$ using eq. \Ref{3.2.121} shown by the bold dot is only slightly shifted up compared {to} that for the exact solution. This tiny difference appeared due to several reasons. First, the value of the critical rotation frequency with the analytical solution corresponds to $\Omt_{cr} =0$, $\Om_{cr} =m/l$, cf. eq. \Ref{3.2.10}, whereas for the exact solution it corresponds to a slightly (for $mR\gg 1$) larger value $\Omt_{cr} =j_{\nu,1}/R$ given by eq. \Ref{3.2.4}, see the insert in figure \ref{fig:1}. Second, the analytical expression for the energy \Ref{3.2.12b} is an expansion in the value $1/(\Omt R)$. Third, the analytical solution \Ref{3.2.12} becomes at $\Omt r\lsim 1$ only an appropriate interpolation satisfying the correct boundary condition.

\begin{figure}
	\includegraphics[width=0.9\textwidth]{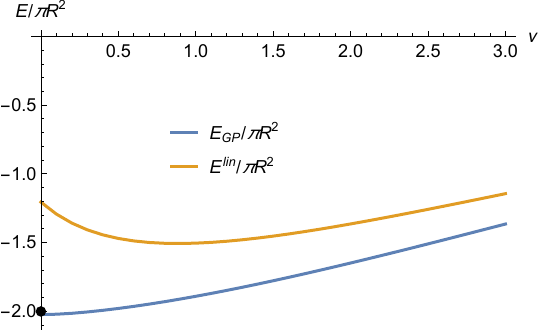} 
	\caption{
		For the thin flux tube, the condensate energies corresponding to numerical solutions of the exact GP-like equation and the linearized equation for $\tilde\Om=1.52$  ($\Om=0.036$ at $l=50$ and $\epp=1.5$) as a function of $\nu$. For the exact solution the minimum of the energy occurs at $\nu=0$, whereas for the linearized solution it occurs at $\nu\approx 1$. The bold dot shows the minimal energy given by \Ref{3.2.12}, {which is} found with the approximate analytical solution \Ref{3.2.121}. 	}		\label{fig:3}	
\end{figure}

{\bf Case $\nu\neq 0$.} For $\nu\neq 0$ the exact solution forms a hole at small $r$. {This} occurs since the appropriate asymptotic solution for $r\to 0$ is $\phi\propto r^\nu$. With increasing $\nu$ the hole is shifted to larger $r$; however, the latter case is energetically less favourable as follows from figure \ref{fig:3}.

An interpolation solution matching the correct asymptotic {behavior} for $y\to 0$ with the approximate solution for $y\gg \nu$ ($r\gg \nu/ \Omt $), {and} satisfying the boundary condition $\chi (y_0=\Omt R)=0$ is as follows
\eq{3.3.A}{ \chi_{int} (y)=F(y) \chi^{(1)}(y)\,,\quad F(y)=\frac{y^{\nu}}{(\nu+y^{2})^{\nu/2}}\,,
}
with $\chi^{(1)}(y)$ from eq. \Ref{3.2.12}. A similar interpolating solution, $\chi_{int} (y)=F(y)$ at $\nu =1$, is employed in the description of the vortices in {He-II}, see \cite{PS2002}.
In dimensional variables it reads
\eq{3.3.C}{ \phi_{int} (r)=\frac{\Omt}{\sqrt{\lambda}}
	\frac{( \Omt r)^{\nu}}{(\nu +(\Omt r)^{2})^{\nu/2}}\mbox{th} \frac{\Omt(R-r)}{\sqrt{2}}\,.
}
In figure \ref{fig:10_33}
we show plots similar to figure 2 but for $\nu =1$. As we see, the approximate numerical solution of the linearized equation fits well the solution of the GP-equation in the vicinity of the critical point whereas for larger $\Om$ the former solution becomes not applicable. {In contrast}, the analytical interpolating solution better fits the solution of the GP-equation for larger $\Om$, when the criterion $\Omt R\gg \nu$ is better fulfilled.

\begin{figure}
	\includegraphics[width=0.9\textwidth]{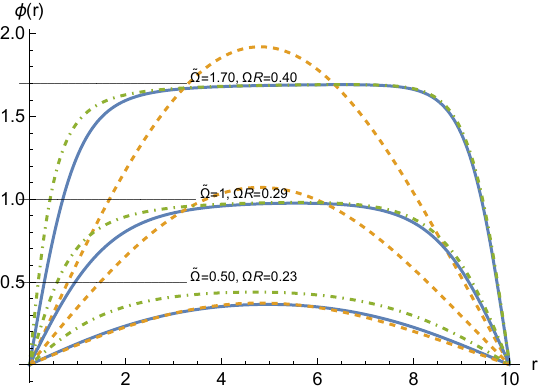} 
	\caption{
		For the thin flux tube, at $\nu=1$, the exact solutions $\phi(r)$ of the GP-like equation (solid lines), the solutions of the linearized equation (dashed lines) and the interpolating solution \Ref{3.3.C} (dot-dashed lines) for several values of the parameter $\Om R$,
		$l=49$, $\delta_\Phi=50$, and $R=10, m=1, \lambda =1$.
	}		\label{fig:10_33}
\end{figure}

{Above we studied the limiting case of the infinitely thin flux tube. However, approximately all the results hold for a thin flux tube of finite size provided the typical scale, $r\sim r^{i}_{typ}$ at which the solution in the interior, at $r<R_s$, changes significantly is much larger than $R_s$. {Under} this condition, instead of solving a complicated problem of finding the interior solution $\phi_i(r)$ and matching it at $r=R_s$ (together with the first derivative) with the exterior solution $\phi_e(r)$, for the case $\nu =0$ one may {set} $\phi_e (R_s)=const$, taking the constant {to be} the same as in the case of the infinitely thin flux tube at $R_s=0$, and for $\nu\neq 0$ setting $\phi_e (R_s)=0$. {An} example showing the efficiency of such a procedure is demonstrated in Appendix \ref{ApA}.
	In the case $r_{typ}^i\lsim R_s$, although correct matching is required to find the solution in the whole region of $r\leq R$, it almost does not affect the value of the energy provided $r_{typ}^e\sim 1/\Omt\gg R_s$, since the integral in \Ref{2.2.7} is then characterized by the distances $r\sim (r_{typ}^e, R)$. }

{In conclusion}, results derived above
for the infinitely thin flux tube approximately hold in the case {of} finite radius $R_s$ {when} $r^{i}_{typ}\gg R_s$.


\subsection{\label{T3.3}Rotation with magnetic field in the whole volume of the cylinder}
In this subsection we consider a homogeneous field that fills the whole cylinder. For this in the relations \Ref{2.3.1}, \Ref{2.3.2} we set $R_s=R$, i.e. now $\mu =\delta_\Phi=B_0 r^2/2$ in the whole volume $r\leq R$.
As before we use \Ref{3.0.1} and study eq. \Ref{3.0.2} with $k_z=0$.

\subsubsection{\label{T3.3.1}Linearized equation}
%

Introducing a new variable in the radial function $\phi(r)$,
\eq{3.3.1}{ \rho=\frac{eB_0 r^2}{2},
}
the solution \Ref{3.0.3}, which is regular at the origin, can be expressed in terms of the Kummer's (confluent hypergeometric) function,
\eq{3.3.2}{ \phi(r)=\rho^{|l|/2}e^{-\rho/2}M(\al,\be,\rho).
}
The parameters are related by
\eq{3.3.4}{ \al = \frac{|l|-l+1}{2} +\frac{-(\Om l)^2+m^2 +\epp^2}{2eB_0},\quad \be =|l|+1.
}
For later {use} we mention the series expansion
\eq{3.3.5}{M(\al,\be,\rho)=1+\frac{\al}{\be}\rho+\frac{\al(\al+1)}{\be(\be+1)2!}
	\rho^2+\dots\,.
}
The Dirichlet boundary condition at $r=R$, which must be imposed on the radial function, results with \Ref{2.3.2} in
\eq{3.3.6}{ M(\al,|l|+1,\delta_\Phi)=0.
}
This is an equation for $\al$ and its solutions $\al_n(l,\delta_\Phi)$ ($n=0,1,...$) define the energies
\eq{3.3.7}{ \epp_{n,l}=\sqrt{(\Om l)^2
		-m^2-k_z^2-eB_0(-2\al_n(l,\delta_\Phi)+|l|-l+1)}.
}
We are interested in the energetically lowest solution and restrict {ourselves} therefore to $l\ge0$.
In the case of a magnetic field homogeneous in the whole space (i.e. in the limit $R\to\infty$), the solutions $\al$ of \Ref{3.3.6} are integer numbers, $\al_n(l,\delta_\Phi)=-n$, $n=0,1,\dots$, and describe the well-known Landau levels, and the Kummer function turns into the Laguerre polynomials. These solutions are degenerate with respect to the orbital momentum $l$ with a degeneracy factor $N=eB_0 R^2/2=\delta_\Phi\geq l$, being the number of states in the interval $L_z\Delta k_z/(2\pi)$.
The energy \Ref{3.3.7} is minimal for $\al_n =0$. In general, the relation $N\geq l$ does not apply for a field defined in a restricted space, e.g. in the case of the boundary condition at $r=R$, see the example shown in figure \ref{fig:5} below. Also in the latter case there is no solution at $\al =0$.

From $\epp=0$, similar to \Ref{3.2.4}, we define the critical rotation frequency $\Om_{cr}$, above which an instability sets in,
\eq{3.3.8}{  \Om_{cr}=\frac{1}{l}\sqrt{m^2+eB_0(-2\al+1)},
}
and represent the ground state energy $\epp_0$ in terms of $\Om_{cr}$,
\eq{3.3.9}{ \epp_0 = l\sqrt{\Om^2-\Om_{cr}^2}\,.
}

In general $\al$ is not necessarily an integer number. Nevertheless, a striking{ly} simple way to get information from the Kummer function is {to terminate} the series expansion \Ref{3.3.5} by taking for $-\al$ a natural number, which was used in \cite{liuz18-120-032001}. In the latter case the minimal energy corresponds to $\al=-1$. If one took $\al=-2$ or smaller, one could find the relations (more than one and non-linear) between the flux and the orbital number, for which the boundary conditions are fulfilled. All these have $\delta_\Phi>l$.
So, taking
$\al=-1$ and $\rho=\be$ in \Ref{3.3.5},
we arrive at a solution of the equation for the boundary condition, which is valid with $\be$ from \Ref{3.3.4} and $\delta_\Phi$ from \Ref{2.3.2} for
\eq{3.3.10}{ \delta_\Phi=l+1, \qquad
	eB_0=\frac{2\delta_\Phi}{R^2}=\frac{2(l+1)}{R^2}.
}
Accordingly, other expressions simplify, for example,
\eq{3.3.11}{ \Om_{cr}=\sqrt{\frac{6(l+1)}{l^2 R^2}+\frac{m^2}{l^2}}
}
follows. Thus, for large $l$ and flux (with \Ref{3.3.10}), the critical frequency goes to zero, $\Om_{cr}\sim \sqrt{6/l}$.
%
%
For the energy parameter $\epp$, \Ref{3.3.9}, we get simply $\epp\sim l \Om$, i.e., it grows linear{ly} with $l$. For the condensate energy \Ref{2.4.8} we need the parameters $a$ and $b$, see \Ref{2.4.5}, for large $l$. For $\al =-1$, the wave function is simply
\eq{3.3.12}{ \phi(r) =
	\rho^{l/2}e^{-\rho/2}\left(1-\frac{r^2}{R^2}\right),
}
and the integrals in \Ref{2.4.5} can be taken explicitly. Their values are presented in Appendix \ref{ApB}.

From eqs. \Ref{3.3.13}, \Ref{3.3.15} of Appendix \ref{ApB}, the expansion for large $l$ follows for the coefficients $a$ and $b$ in \Ref{2.4.5} and the mean radius from the Stirling approximation (and some generalization),
\eq{3.3.16}{\frac{a^2}{b}\simeq \frac{2\sqrt{\pi}R^2}{3\sqrt{l}},\qquad
	\langle r\rangle/R \simeq 1-\sqrt{\frac{2}{\pi l}}.
}
Finally, for the ground state condensate energy per unit length \Ref{2.4.8}, in this approximation we arrive at
\eq{3.3.17}{ E_{0}^{lin}\simeq
	-\frac{4\sqrt{\pi}R^2\Om^4 l^{7/2}}{3\la}\,, \quad l=\delta_\Phi -1\,.
}
If {this} were so for an arbitrary relation between $l$ and $\delta_\Phi$ and in the self-consistent treatment of the fields $\phi$ and $\vec{A}$, it could cause a problem. Indeed, then the energy would not be bounded from below since \Ref{3.3.17} for large $l$ cannot
be balanced by the classical electromagnetic energy term \Ref{2.3.2}, which grows		quadratically with $l$. However, in the given paper we treat the field $B$ as an external one and in \Ref{3.3.17} we have $l=\delta_\Phi -1$, and the values of $B_0$, $R$ and $\Omega$ are in our case the external input parameters. In such a case the problem does not occur.

\subsubsection{\label{3.3.3}Numerical solutions of GP-like equation}

\begin{figure}[t]
	\includegraphics[width=0.9\textwidth]{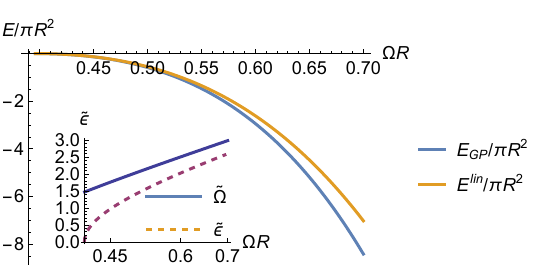} 
	\caption{
		For the magnetic field in the whole cylinder, the condensate energy density, $\frac{E^{GP}}{\pi R^2}$, see \Ref{2.2.8}, and $\frac{E^{lin}}{\pi R^2}$, see \Ref{2.4.8}, as functions of $\Om R$. The parameters are $m=1$, $R=10$, $\delta_\phi=50$ and $l=45$.
		The inset shows the dependence of $\Omt$ and $\epp$ on $\Om R$.
	}		\label{fig:3a}	
\end{figure}
We calculated  numerical solutions of the full non-linear equation \Ref{3.0.2}, again with $k_z=0$. The substitution \Ref{3.2.8} turns this equation into
\eq{3.3.30}{ \left(  -\frac{\pa^2}{\pa\, r^2}-\frac{1}{r}\frac{\pa}{\pa\, r}
	+\left(\frac{l-\delta_\Phi r^2/R^2}{r}\right)^2-\Omt^2+\lambda \phi^2(r) \right)\phi(r)=0,\quad \phi(0)=0,\ \ \phi(R)=0,
}
again with $\Omt$ given by eq. \Ref{3.2.10}. Now we employ the condition $\phi(0)=0$ since the asymptotic {behavior} of the regularly behaving solution of \Ref{3.3.30} at $r\to 0$ is $\phi\propto r^l$ for $l\neq 0$, as for the linear equation.
%
{T}he numerical solution of  eq. \Ref{3.3.30} was found {by} applying Newton's iteration method to a discretization of the equation with a division of the interval into $10^5$ points. The program {package} NINE from JINRLIB \footnote{

Figure \ref{fig:3a} demonstrates energy densities $\frac{E_{GP}}{\pi R^2} $ and $\frac{E^{lin}}{\pi R^2}$ as functions of $\Om R$, the same as in figure \ref{fig:1} but now for the uniform magnetic field in the whole cylinder.
It is seen, that as it should be,  $E^{lin}>E_{GP}$.
Moreover, we see that in the case of   the magnetic field in the whole cylinder,  the energy is higher than for the thin flux tube, both for solutions of the linearized and exact GP-like equations. See also figure \ref{fig:8}{} below.  The first excited state  solution of the linearized equation appears  for considerably larger  $\Om_{cr}^{(1)} \approx 0.059$ than the value $\Om_{cr}\approx 0.039$ in this case.
The {inset}  shows the relation {between} $\epp$ and $\Omt$ as functions of $\Om R$. Here we see that these values differ significantly, whereas in the case of the thin flux tube they almost coincided everywhere except {in the} vicinity of the critical point. The difference appears due to the difference in the presentation of the centrifugal terms in eqs. \Ref{3.2.1} and \Ref{3.3.30}.

\begin{figure}[t]
	\includegraphics[width=0.75\textwidth]{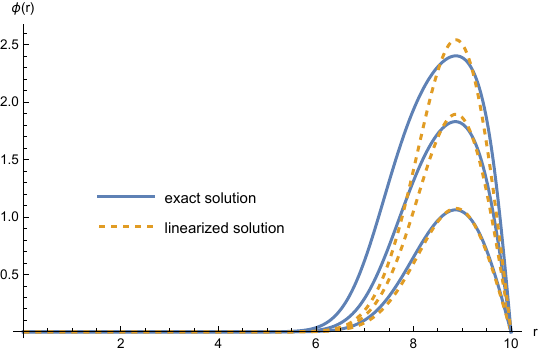} 
	\caption{ 
		The case of the magnetic field in the whole cylinder $r\leq R$.
		The solutions $\phi(r)$  of the exact equation \Ref{3.3.30} (solid lines) and of the linearized equation, \Ref{3.0.3}, (dashed lines) for $l=45$, $\delta_\Phi=50$ and $\Om R=0.45,0.53,0.62$,   correspondingly $\epp=0.07,{1.33,1.97} $, from bottom to top.
		Other parameters are chosen  the same as in previous figures.
	}		\label{fig:4}	
\end{figure}

\begin{figure}[h]
	\includegraphics[width=0.75\textwidth]{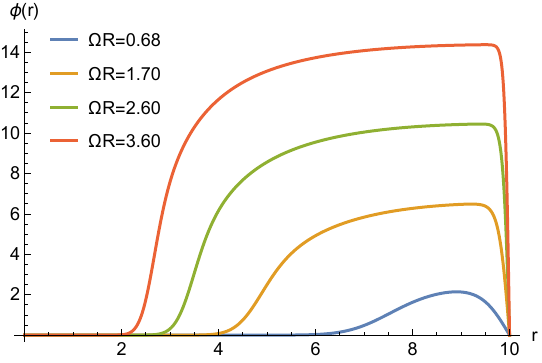} 
	\caption{
		Solutions $\phi(r)$ of the exact {equation} \Ref{3.3.30} for rather large  values of $\Om R= 0.68,1.7,2.6,3.6$ (from bottom to top), demonstrating the formation of a plateau. These values, except the first one, are outside the causality constraint. The parameters are $R=10$, $l=40$, $\delta_\phi=50$.
	}		\label{fig:6}	
\end{figure}

\begin{figure}[t]
	\includegraphics[width=0.45\textwidth]{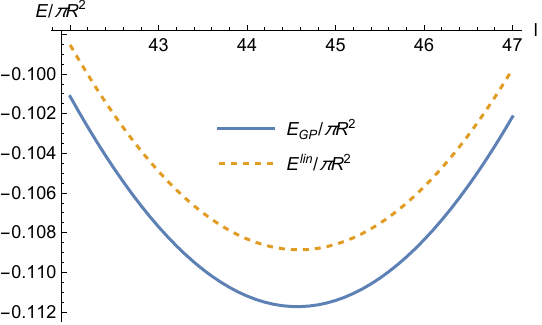} 
	\includegraphics[width=0.45\textwidth]{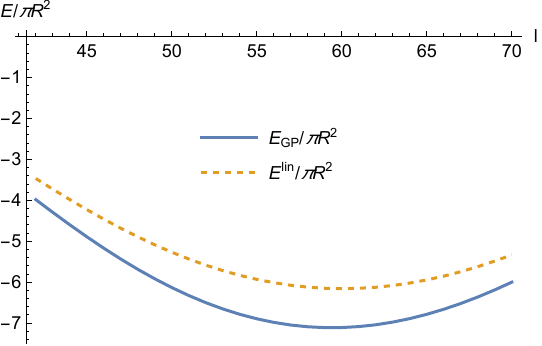}
	\caption{
		{The} case of the magnetic field in the whole cylinder. The condensate energy densities, $\frac{E^{{GP}}}{\pi R^2}$, of the full equation \Ref{2.2.8} (solid line), and  $\frac{E^{lin}}{\pi R^2}$ (dashed line),  of the linearized equation \Ref{3.0.3}, as functions of $l$ for   $\delta_\Phi=50$ with $\Om R=0.45$ (left panel) and $\Om R=0.65$ (right panel).  The minima
		are located at $l=44.6$ and the energy values are	
		$\frac{E_{GP}}{\pi R^2}=-0.112$,
		$\frac{E^{lin}}{\pi R^2}=-0.109$, in the left panel and
		$\frac{E_{GP}}{\pi R^2}=-7.11$ at $l=59$, $\frac{E^{lin}}{\pi R^2}=-6.16$ at $l=60$, in the right panel.
	}		\label{fig:5}	
\end{figure}

\begin{figure}[h]
	\includegraphics[width=0.8\textwidth]{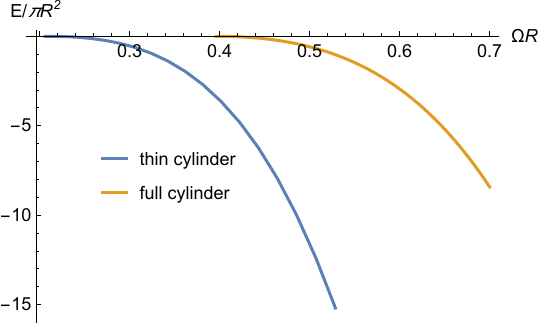} 
	\caption{
		The condensate energies for the thin flux tube and for the magnetic field in the whole cylinder as {functions} of $\Om R$.  The parameters are $m=1$, $R=10$, $\delta_\phi=50$, and $l=50$ for the thin flux tube and $l=45$ for the field in the whole volume (giving the energy a minimum in the corresponding cases).
	}		\label{fig:8}	
\end{figure}
	
In figure \ref{fig:4}, the solutions of the linearized and the exact equations are shown for $l\neq \delta_\Phi$ at $\Om R<1$. In both cases the solutions are rather close {to} one another, more {so} for smaller $\Om R$, $\Omt$, $\epp$, as expected. {The} solutions have a Gaussian-like form and the maximum of $\phi$ and mean radius are shifted to $r$ closer to $R$ and the plateau which we observed in the case of a thin flux tube is not formed. {This is because}
the asymptotic solution at $r\to 0$ behaves  as $\phi\propto r^l$ and here $l \gg 1$. In the case of the thin flux tube we had $\phi\propto r^\nu$ and considered  not large values of $\nu$ since the minimal energy corresponds then to $\nu =0$. 
In figure 7  we demonstrate that a plateau would be  nevertheless  formed also in the case of the field in the full cylinder,   however not in the causal region, but beyond, in the region $\Om R>1$, which is considered unphysical in the context of this paper.


The dependence of the condensate energy on the orbital momentum is shown in figure \ref{fig:5} for the exact and the linearized solutions. {Figure} \ref{fig:5} shows that for the fixed input parameters there exists a fixed optimal value of $l$ corresponding to the minimum of the energy in both cases of solutions of linearized and exact GP-like equations. The minimum of the energy is shifted to lower orbital momenta compared to $\delta_\Phi$ ($l_m-\delta_\Phi<0$) for rather   small $\Om$ {(for $\Om R=0.45$ in the example shown in the figure,  $\Om_{cr}R\simeq 0.395$)}
and towards {larger} $l$ ($l_m-\delta_\Phi>0$) for larger $\Om R$ (for $\Om R=0.{65}$ in the example shown in the figure). This is {opposite to} the case of the thin flux tube where $\nu$ corresponding to the minimum of the energy, $\nu_m$, enters only as $\nu_m^2$.
	
In figure \ref{fig:8} we compare energy densities of the condensate as functions of $\Om R$ at a fixed value of the flux $\delta_\Phi$, for the field in the thin flux tube and in the whole cylinder.  As {can be} seen, the energy density of the thin flux tube is lower than that for the field occupying the whole cylinder. Although in our artificial problem the infinitely thin flux tube is energetically favorable,  in a more realistic consideration the minimal size of the flux tube is determined by the minimal length scale, i.e., by the minimal value of $r_i$ and $r_e$ as it was mentioned.
Recall that in the case of superconductors there appear two characteristic length scales,  the penetration depth $d$ of the magnetic field and the coherence length $\xi$ characterizing the order parameter. In the case of superconductors of the second kind the ratio of these lengths, the Ginzburg-Landau parameter, is $\kappa =d/\xi>1/\sqrt{2}$. Then  the condensate vortices form the Abrikosov lattice with $l=1$ angular momentum for each vortex \cite{LP1981}.
Thus we may conjecture that in our case, {in} a  self-consistent treatment of the rotating cylinder in the applied magnetic field, it  will be confined in {flux} tubes. If {this} were so, it could raise a question about formation of a lattice of such {flux} tubes similarly to that {which} occurs in superconductors \cite{LP1981}, but in our case with a  large $l$  for each flux tube, see a discussion in \cite{vosk25-111-036022}.	

\section{\label{T4}Conclusion}
In this paper we studied the problem of the instability of the vacuum {with} respect to the production of a charged scalar (or pseudoscalar) field under the rapid rigid rotation in the presence of an external
{magnetic} background field. Of practical interest is the problem for the pion field in application to heavy-ion collisions. Therefore, for numerical estimates we used values of parameters relevant to this case{;} for instance, we employed that $mR\gg 1$, $B\lsim m^2_\pi$ and $\Om\lsim 0.1 m_\pi$.

We started with the derivation of the Gross-Pitaevskii-like equation for the charged scalar (pseudoscalar) mean field in the uniformly rotating frame in the presence of the external electromagnetic background field introduced initially in the resting frame.
{We explicitly performed the transition from variables in the resting frame to the {rotating} frame used in the sense of a passive transform. We mention that for the charged scalar field the Klein-Gordon equation in the {rotating} frame is the same as in the local flat frame, which is also {used} for {this} kind of problem{s}. } The causality condition requires that $\Om R<1$, {which} is ensured by imposing the Dirichlet boundary condition at $r=R$.
It raises the question {of} to {what} extent this circumstance influences the dynamics of the scalar field. As {can be} seen in figure \ref{fig:4}, where we showed the behavior of the scalar field in the {rotating} frame in the presence of the external uniform magnetic field, the condensate field is concentrated near the surface of the ``mathematical'' cylinder, which implies a large influence. Thereby we may conclude that passive rotation of the frame and active rotation of the empty ``physical'' cylinder with a {non}-transparent wall represent, in a sense, different cases. Although in the former case interpretation causes questions, the latter case allows for a {clearer} physical interpretation. Thereby we focused on the latter case. Thus we related the metric to the rotating empty cylinder and considered the case when the flux tube (solenoid) of radius $R_s\leq R$ generating the uniform magnetic field for $r<R_s$, see eq. \Ref{2.3.1}, is {at} rest in the frame associated with the rotating cylinder of radius $R$ ({rotating} frame). The aim was to describe the response of the scalar field vacuum {to} such a magnetic field in the {rotating} frame.}

In the {foregoing} section, to be specific, we considered two configurations of the magnetic field{:} an infinitely thin flux tube (the solenoid of radius $R_s\ll R$ generating uniform magnetic field inside it) and a field homogeneous in the whole cylinder (the solenoid of radius $R$). {The} case of the uniform magnetic field in the whole volume of the rotating cylinder was previously studied in \cite{liuz18-120-032001,guot22-106-094010,vosk25-111-036022}. {The} case of a thin flux tube in the {rotating} frame was not studied {before}, to the best of our knowledge.
In both cases we found a numerical solution of the linearized equation, as an approximation, and a numerical solution of the full non-linear Gross-Pitaevskii-like (GP-like) equation, as well as {found} approximate analytical solutions, where {possible}. For the solution of the linearized equation, in the case of the infinitely thin flux tube, the condensate energy is expressed in terms of {zeros} of the Bessel function. For the case of the uniform magnetic field in the cylinder one needs to find a zero of the Kummer function, which problem is more delicate, especially for large values of parameters like the rotation frequency. However, for some special relations between magnetic flux and orbital momentum, e.g., when the magnetic flux $\delta_\Phi$ and the orbital momentum $l$ are connected as $\delta_\Phi =l+1$, an explicit solution exists. In the numerical approach, we adopted a relaxation method and a 4th order Runge-Kutta approach. Moreover, for the thin flux tube in the case $\delta_\Phi =l$ we found the approximate analytical solution of the GP-like equation. The latter case is of special interest, since we showed that the energy of the produced
{condensate takes a minimum precisely for $l=\delta_\phi$}, see figure \ref{fig:3}.

In the case of the flux tube, due to the interplay between rotation and magnetic background field, after the rotation frequency reaches the critical value, given by eq. \Ref{3.2.4} for $\delta_\Phi =l$,
the scalar charged field can be produced from the vacuum in the {rotating} frame. This instability results in the formation of the classical field (the condensate). We calculated its energy in a range of parameters, which may be of relevance in heavy-ion collisions for pions.
We demonstrated that the exact and the linearized solutions are very close in the vicinity of the critical point but begin {to} deviate essentially for a larger $\Om$, see figures \ref{fig:1} and \ref{fig:2}. They show at the threshold the behavior typical {of} the second-order phase transition. The approximate analytical solution {coincides well} with the numerical solution of the GP-like equation except {in} the vicinity of the critical point.
We noted that the critical point obtained with the approximate analytical solution for the flux tube at $\delta_\Phi =l$ is slightly shifted to a smaller value (shown by the bold dot in figure \ref{fig:3}), whereas the value of the correction is $1/(\Omt R)$, where $\Omt =\sqrt{\Om^2-m^2}$.

For the exact and approximate analytical solutions of the GP-like equation the minimum of the energy appears precisely at $l=\delta_\phi$, whereas for the case of the linearized equation it is {slightly} shifted to $\nu\approx 1$, see figure \ref{fig:3}. Due to the dependence only on $\nu^2$ the possibilities $l<\delta_\Phi$ and $l>\delta_\Phi$ are indistinguishable in this case.

Figure \ref{fig:2} shows that with increasing $\Om R$ the plateau solution is formed in the case of the flux tube at $l=\delta_\Phi$, which is well described by the approximate analytical solution \Ref{3.2.121}. The mean radius of the exact solution corresponds to $r\sim 0.5 R$ accordingly.

We argued, {under} which conditions a thin flux tube can be considered as an infinitely thin one.
In the case $\delta_\Phi\neq l$ for the thin flux tube we found an interpolating analytical solution between two asymptotic regimes, for small $r\ll \nu/\Omt$ and large $r\gg \nu/\Omt$, see eq. \Ref{3.3.C}. It demonstrates the appearance of a hole starting at small $r$, which origin is associated with the appearance of the centrifugal barrier for $\nu\neq 0$. In figure \ref{fig:10_33} we demonstrated that for $\nu\sim 1$ the numerical solution of the exact GP-like equation and the analytical interpolation expression {coincide well} except {in the} vicinity of the critical point, whereas the solutions of the exact GP-like equation and the linearized one {coincide well} in the vicinity of the critical point.

For the uniform magnetic field in the whole cylinder our results obtained within the linearized approach are in general agreement with \cite{guot22-106-094010}, {with regard to} the critical frequency, the statement that excited states are occupied only at $\Omega$ essentially larger than $\Om_{cr}$ and that the minimum of the condensate energy occurs at a finite value of the orbital moment $l$, see figure \ref{fig:5}.
However, in \cite{liuz18-120-032001,guot22-106-094010,vosk25-111-036022} the authors restricted the orbital moments by the number of states (per unit area) $N=\delta_\phi$. This is correct, if {one considers} the system in uniform magnetic field in the unbounded space, whereas the cylinder with $r<R$ corresponds to a bounded space and in this case the states with any orbital momentum $l$ can exist. In the case of the field in the full cylinder it proved to be that the minimum of the energy corresponds to $l<N$ for $\Om$ varied in a vicinity of $\Om_{cr}$ but it corresponds to $l >N$ for a larger $\Om$, see figure \ref{fig:5}.

The easiest approximation method for the GP-equation is its linearization. As shown in \cite{bord25-40-2543018}, {this} solution can be viewed as the first step in a perturbative expansion for small $\epp$ and we confirmed that for small $\epp$ the solutions of the linearized and the full equations are very close. In general, the linearized equation has a full spectrum of modes with higher $\epp$. These modes were used in \cite{guot22-106-094010} to expand the full solution. We confirm their finding that for small $\epp$ only the lowest modes need to be included by our observation of the mentioned closeness.

We used the analytical solutions of the linearized equation at $\delta_\Phi =l+1$. The value of the critical angular velocity \Ref{3.3.11} proved to be essentially larger than in the case of the flux tube at the same value of the magnetic flux.

In the case of the field in the full cylinder, figure \ref{fig:3a} shows energies found with numerical solutions of the GP-like equation and the linearized equation. The solution demonstrates the occurrence of the second order phase transition at $\Om=\Om_{cr}$. In the vicinity of the critical point the solution found in the linearization method
coincides well with the numerical solution of the GP-like equation, see figure \ref{fig:4}.
With increasing $\Om$ the linearization procedure becomes less appropriate. The solutions demonstrate a big hole up to $r\lsim R/2$. Here the hole is larger than in the case of the thin flux tube since in the former case the centrifugal barrier is much larger.
Notice however that although in the case of the field in the cylinder the plateau does not arise, as is shown in figure \ref{fig:4} at $\Om R<1$, it would reappear for $\Om R>1$, as {shown} in figure \ref{fig:6}.

Finally, as {has} been mentioned, we demonstrated that the energy in the case of the thin flux tube is lower than that for the magnetic field in the whole cylinder,
see figure \ref{fig:8}, both for solutions of the linearized and exact GP-like equations. Thereby, {although} we considered the magnetic field as an external one, we may conjecture that in the self-consistent problem the uniform magnetic field applied to the rotating cylinder will be confined in a narrow flux tube. If so, it will then raise a question about the formation of the lattice
of flux tubes similarly to that which {occurs} for {vortices} in superconductors of the second kind \cite{LP1981}, but in our case with a large flux in each flux tube.

In the present paper we ignored the back reaction of the produced charged condensate field on the distribution of the electric and magnetic fields inside the cylinder. With increasing $\Om$ significantly above $\Om_{cr}$ the inclusion of such a response can become very important. In a somewhat similar problem of the behavior of the superconductors of the second {type} in the external uniform magnetic field, a back response of the order parameter on the distribution of the magnetic field proves to be rather weak only near the critical values of the magnetic field $H_{cr,1}$ and $H_{cr,2}$. This problem needs further investigation.

Notice that in the case of the non-abelian gluon fields the vacuum placed in the uniform magnetic field may consist of the lattice of the Nielsen-Olesen vortices. We hope that our results can be helpful {in studying} this problem in the {rotating} frame. Further investigation of the Casimir effect under the rotation can be also of interest \cite{bord25-40-2543018}. Under the rapid rotation of a piece of the nuclear matter the inhomogeneous $\pi^0$ and $\pi^{\pm}$ condensates can be produced via the {anomalous} Wess-Zumino-Witten contribution to the Gibbs free-energy density \cite{Yamamoto}. In the case of extremely rapidly rotating nuclear drops, the occurrence of a charged pion vortex condensate may stabilize them in the rotating frame \cite{vosk24-109-034030,vosk25-111-036022}. In peripheral heavy-ion collisions, the presence of the strong magnetic and {rotational} fields may stimulate the formation of the pion condensate in the {rotating} frame \cite{vosk25-111-036022}. All these problems also need {further} study.

\section*{Acknowledgements}
The authors acknowledge valuable assistance from the Laboratory of Information Technologies of JINR, Dubna, provided by J. 
Bu$\check{\rm s}$a and E. Zemlyanaya. We also thank H. Grigorian, E. E. Kolomeitsev and I. G. Pirozhenko for fruitful discussions.

\appendix
\section{Simple formulas for the solution of
	equation \Ref{3.0.3}
}\label{ApA}
Here we display the simple formulas for the solution of \Ref{3.0.3}, i.e. the linearized eq. \Ref{3.0.2}, with the background field \Ref{2.3.1}.
%
We represent the solution as consisting of two parts{:} $u_i(r)$ for the interior part, i.e., for $0\le r\le R_s$, and $u_e(r)$ for the exterior part, i.e., for $R_s\le r\le R$,
\eq{A.2}{\phi(r)=u_i(r)\, \Theta(R_s-r)+u_e(r)\, \Theta(r-R_s),
}
supplemented by the continuity condition for the function and its derivative at $r=R_s$. With the notation $\rho=\delta_\Phi \frac{r^2}{R_s^2} $, the interior solution is
\eq{A.3}{ u_i(r) & = \rho^{|l|/2}e^{-\rho/2}M(\al ,|l|+1,\rho),
}
where $M(\al,|l|+1,\rho)$ is the Kummer{'}s function (confluent hypergeometric function) and
\eq{A.4}{ u_e(r) & = C_1 \,J_\nu(\Lambda r)+C_2\, Y_\nu(\Lambda r),
}
which is a combination of the Bessel functions. The coefficients are
\eq{A.5}{ C_1 &= \frac{\pi R_s}{2}\left( u_i(R_s) Y'_\nu(\Lambda R_s)- u_i'(R_s) Y_\nu(\Lambda R_s) \right),
	\nonumber	\\
	C_2 &=\frac{\pi R_s}{2}\left(- u_i(R_s) J'_\nu(\Lambda R_s)+ u_i'(R_s) J_\nu(\Lambda R_s) \right),
}
with parameters
\eq{A.6}{\nu=|\delta_\Phi -l|,\quad \al =\frac{1}{2B_0}\left( m^2+eB_0\frac{|l|-l}{2}-(\epp-\Om l)^2\right),
	\quad \Lambda=\sqrt{\epp^2+(\Om l)^2-m^2}.
}
The boundary condition at $r=0$ is fulfilled by the choice of the interior function and the boundary condition $\phi(R)=0$ is an equation for the energy $\epp$.

We mention that for $R_s\to 0$, the coefficients behave as
\eq{A.7}{ C_1\sim R_s^{-\nu},\quad C_2\sim R_s^\nu,
}
and in the limit only the function $J_\nu(\Lambda r)$ remains and the wave function at $r=R_s\to0$ tends to zero (case $\nu\ne0$).

The flux tube of a finite radius $R_s$ can be approximately described in the limit $R_s\to 0$ provided the typical scale, at which the solution $u_{i} (r)$ changes essentially, $r_{typ}^i\sim \sqrt{l/(eB_0)}$, satisfies the inequality $r_{typ}^i\gg R_s$. Then, whether we match the interior and exterior solutions and their derivatives properly at $r=R_s$ or take as the boundary condition simply $u_e (R_s)\approx 0$ {hardly} matters. {An} illustration is performed in figure \ref{figA} for $R_s =0.5/m$. We found the solutions $u_i$ and $u_e$ and matched them (and the first derivatives) at $r=R_s$. However, in this case $r_{typ}^i\simeq 6.7/m\gg R_s$ and one may see that $u_e (r=R_s)$ is already tiny. Thereby instead of finding the solution $u_i(r)$ and performing the matching procedure we can restrict ourselves to finding only $u_e (r)$ taking the boundary condition $u_e (R_s)\simeq 0$.

\begin{figure}[H]
	\includegraphics[width=0.6\textwidth]{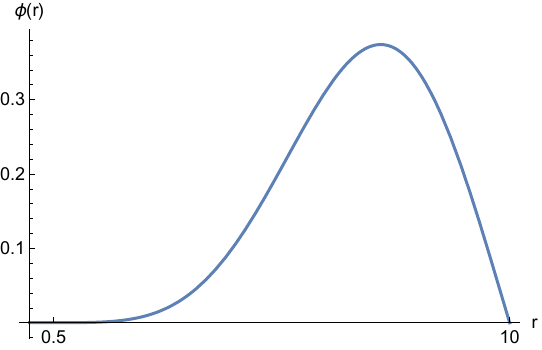}
	\caption{
		The solution $\phi(r)$ of eq. \Ref{3.0.3} for $m=1$, $l=45$, $\delta_\Phi=50$, $\Om=0.08$, $R=10$, $R_s=0.5$. The eigenvalue in equation \Ref{3.0.3} is $\epp=3.35$. The tick shows $R_s$. }		\label{figA}	
\end{figure}

\section{Mean radius for the solution of linearized equation \Ref{3.0.2}}\label{ApB}
With the field in the form \Ref{3.3.12}, the calculation of the coefficients $a$ and $b$ in \Ref{2.4.5}, entering the expression for the energy $E^{lin}_{GP}$, yields
\eq{3.3.13}{\frac{a^2}{b}=
	\frac{4^{l+2} (2 l+1) (2 l+3)  \left(e^{l+1} (\Gamma (l+3)-\Gamma (l+3,l+1))+(l+1)^{l+2}\right)^2 R^2}
	{(l+1) (l+2) \left(e^{2 l+2} (3 l+2) (\Gamma (2 l+5)-\Gamma (2 l+5,2 (l+1)))+4^{l+3} (2 l+3)
		(l+1)^{2 l+4}\right)},
}
in terms of generalized Gamma functions. For the mean radius, we obtain
\eq{3.3.15}{\langle r\rangle =
	\frac{(l+2)  \left(e^{l+1} (4 l+7) \left(\Gamma \left(l+\frac{7}{2}\right)-\Gamma
		\left(l+\frac{7}{2},l+1\right)\right)+2 (l+1)^{l+\frac{7}{2}}\right)R}{\sqrt{l+1} (2 l+3) (2
		l+5) \left(e^{l+1} (\Gamma (l+3)-\Gamma (l+3,l+1))+(l+1)^{l+2}\right)}.
}
The asymptotics of these expressions for large $l$ are given in the text by eqs. \Ref{3.3.16} and \Ref{3.3.17}.

\end{document}